\documentclass{emulateapj}

\lefthead{van der Wel et al.}
\righthead{Axis ratios of Galaxies and the Morphology-Density Relation}
\slugcomment{The Astrophysical Journal, Received 2009 November 6; accepted 2010 April 2}
\begin{document}

\newcommand{\kms}{\>{\rm km}\,{\rm s}^{-1}}
\newcommand{\reff}{R_{\rm{eff}}}
\newcommand{\msol}{M_{\odot}}

\title{The Physical Origins of The Morphology-Density Relation:\\
  Evidence for Gas Stripping from the Sloan Digital Sky Survey}



\author{Arjen van der Wel\altaffilmark{1}, Eric
  F.~Bell\altaffilmark{1,2}, Bradford P.~Holden\altaffilmark{3}, Ramin
  A.~Skibba \altaffilmark{1,4} \& Hans-Walter Rix\altaffilmark{1}}

\altaffiltext{1}{Max-Planck Institute for Astronomy, K\"onigstuhl 17,
  D-69117, Heidelberg, Germany}

\altaffiltext{2}{Department of Astronomy, University of Michigan, 500
  Church Street, Ann Arbor, MI, 48109, USA}

\altaffiltext{3}{University of California Observatories/Lick
  Observatory, University of California, Santa Cruz, CA, 95064, USA}

\altaffiltext{4}{Steward Observatory, University of Arizona, 933 North
  Cherry Avenue, Tucson, AZ 85721, USA}

\begin{abstract}
  We provide a physical interpretation and explanation of the
  morphology-density relation for galaxies, drawing on stellar masses,
  star formation rates, axis ratios and group halo masses from the
  Sloan Digital Sky Survey.  We first re-cast the classical
  morphology-density relation in more quantitative terms, using low
  star formation rate (quiescence) as a proxy for early-type
  morphology and dark matter halo mass from a group catalog as a proxy
  for environmental density: for galaxies of a given stellar mass the
  quiescent fraction is found to increase with increasing dark matter
  halo mass.  Our novel result is that - at a given stellar mass -
  quiescent galaxies are significantly flatter in dense environments,
  implying a higher fraction of disk galaxies.  Supposing that the
  denser environments differ simply by a higher incidence of quiescent
  disk galaxies that are structurally similar to star-forming disk
  galaxies of similar mass, explains simultaneously and quantitatively
  these quiescence -nvironment and shape-environment relations.  Our
  findings add considerable weight to the slow removal of gas as the
  main physical driver of the morphology-density relation, at the
  expense of other explanations.
\end{abstract}


\section{Introduction}\label{intro}
It has been known for many decades that galaxy morphology and
environment are correlated \citep{smith35, zwicky42, sandage61}.  In
dense environments the fractions of elliptical and S0 galaxies are
higher than in low-density environments, at the expense of a decreased
fraction of spiral and irregular galaxies \citep{dressler80, vogt04b}.
This trend is universal and persists over a large dynamic range in
density \citep{postman84}.  S0 and spiral galaxies have in common that
they have rotationally supported stellar disks.  This suggests that
there may be a direct evolutionary link, in the sense that S0 galaxies
could be the descendants of spiral galaxies that had their
star-formation activity truncated.

The existence of the morphology-density relation (MDR) suggests that
this process may be related to the environment \citep[see, e.g.,][for
an overview of the field]{boselli06}.  Such a process generally
involves the partial or entire removal of the gaseous interstellar
medium from galaxies that become satellites in larger dark matter
halos, through interaction with the intergalactic medium.  Fast,
rather violent stripping of a galaxies' entire interstellar medium is
often considered a viable option \citep{gunn72, quilis00}, but also
more gentle stripping of hot gas outside the cold disk
\citep{larson80, bekki02, mccarthy08}, often called 'starvation' or
'strangulation', and the gradual stripping of neutral gas from the
outer parts of disks \citep[e.g.,][]{chung07} are commonly invoked.
In any case, gas-deficient spiral galaxies in clusters do exist
\citep[e.g.,][]{giovanelli85}, which indicates that gas stripping
occurs.  Eventually, in all these scenarios, gas stripping removes the
fuel for star formation, presumably producing a quiescent galaxy which
is similar to a normal spiral galaxy in terms of its structural
properties such as its bulge-to-disk ratio.  Early on,
\citet{vandenbergh76} described such gas-free, 'amenic' spirals, and
\citet{balogh98} observed how cluster galaxies have lower star
formation rates than field galaxies with the same bulge-to-disk ratio.


However, if S0s occur when a spiral galaxy is stripped from its gas by
interaction with the intragalactic medium, how do we explain the
ubiquitous presence of S0s outside dense environment?  Moreover, over
the years, it has become clear that S0 galaxies differ from the spiral
galaxies in several important ways, implying that the S0 population as
a whole is not simply a population of disk galaxies with little or no
star formation.  S0 galaxies have more pronounced thick disks than
spiral galaxies, as was first shown by \citet{burstein79d}.
Furthermore, the bar fraction among S0s is significantly smaller than
that among spiral galaxies \citep{aguerri09, laurikainen09}.  Perhaps
the most important difference is that S0s have larger bulge-to-disk
ratios than spiral galaxies \citep{dressler80}.  This is supported by
increasingly convincing evidence that the \citet{tully77} relation for
S0 galaxies differs from that of spiral galaxies \citep[][,
M. Williams et al., in prep.]{neistein99, hinz03, bedregal06}.
Although some of these differences may be the result of secular
evolution and evolving stellar populations after the truncation of
star formation, difficulties remain \citep[see, e.g.,][for a
discussion]{dressler84}.  In particular, the large bulges of S0
galaxies cannot be explained.

Besides the disconnect between S0s and spirals, the increased fraction
of elliptical galaxies in dense environments \citep{dressler80} also
challenges the idea that gas stripping explains the MDR.  This
concern, as well as the different properties of spirals and S0s, could
be accommodated by a invoking a process that affects both the structure
and gas content of a galaxy.  Tidal interactions with other galaxies
or the potential of a large halo can strip an in-falling galaxy from
gas, and reduce the stellar disk as well \citep{moore96}.

There are two reasons why it is surprisingly difficult to interpret
the MDR, constrain its origin, and understand its implications for
galaxy formation and evolution in general.  First, 'morphology' is a
phenomenological parameter that is the combination of several physical
quantities (structure, that is, bulge-to-disk ratio or concentration,
and star formation activity). Second, many galaxy properties depend on
one another, and some of these dependencies are much stronger than the
dependency of morphology on environment.  More specifically, although
structure and star formation activity depend strongly on one another
\citep[e.g.,][]{kauffmann03b, hogg04, kauffmann06}, these two
parameters behave distinctly different as a function of environment.
\citet{kauffmann04} and \citet{blanton05b} showed that structure
depends only weakly on environment, whereas star formation activity,
usually as traced by color, decreases significantly from low- to
high-density environments \citep[e.g.,][]{lewis02, hogg03, gomez03,
  hogg04, balogh04, blanton05b, baldry06, skibba09a}.  Moreover,
\citet{balogh98} showed that galaxies that are similar in structure
have lower star formation rates if they are situated in a cluster.
The net effect of these correlations is that the morphological mix
changes with environment \citep{vanderwel08a, bamford09, skibba09c}.

To summarize the above, we have an apparent contradiction: on the one
hand, the structural properties of galaxies do not show strong
environmental dependencies; on the other hand, the relative
elliptical, S0 and spiral galaxy fractions clearly do vary with
environmental density, and these different types of galaxies have
different structural properties.  In this paper, we address this
issue.  We explicitly address the question whether gas stripping can
explain the MDR, even in the face of the evidence that S0s and spiral
galaxies are structurally different, that S0s occur in both low- and
high-density environments, and that elliptical galaxies also prefer
dense environments.  Perhaps contrary to expectation, the listed
evidence does not rule out gas stripping as an explanation for the
MDR.  One should distinguish between the S0 population as a whole, and
those galaxies that have been affected by environmental processes,
giving rise to the MDR.  In other words, the origin of S0 galaxies and
the origin of the MDR are not necessarily the same.  This distinction,
which was already noted by \citet{postman84}, will prove to be
critical.


In this paper, we examine the shapes of a large sample of galaxies
with low star formation activity (quiescent galaxies), as inferred
from their spectroscopic properties, in different environments. The
sample selection from the Sloan Digital Sky Survey (SDSS) is described
in Section \ref{sample}.  The shape parameter we use is the projected
axis ratio. For an individual galaxy this may not contain much
information, but for large samples it becomes a powerful diagnostic.
We take advantage of the uniform and well-calibrated data set provided
by the SDSS, which is ideally suited to disentangle projection effects
and the intrinsic shape distribution of galaxies, as has previously
been demonstrated by \citet{vincent05} and \citet{padilla08}.  Such an
approach is complementary to studies of morphology, defined either
visually or otherwise, because it makes no assumptions about the
connection with galaxy structure.  Because, in addition, we select our
sample spectroscopically, our analysis is completely independent of
the structural properties of different types of galaxies.

The first goal of this paper is to confirm that the decreased star
formation rate in dense environments is reflected as an increased
fraction of quiescent galaxies in massive dark matter halos (Section
\ref{secmdr}).  A halo-based description of environment is more
physical than estimates of the local surface number density of
galaxies, and, moreover, the correlation between galaxy color and halo
mass, and not galaxy density, has been demonstrated to be the
principal driving factor behind the observed correlation between
galaxy color and environment in general \citep{blanton07b}.

The second goal is to determine whether or not gas stripping can
account for the dependence of star formation activity on group mass.
We analyze the axis ratio distribution of quiescent galaxies and its
dependence on halo mass, in Sections \ref{disks} and \ref{mdrdisks}.
Gas stripping of spiral galaxies, which presumably leaves their
structural properties mostly unchanged, and alternative scenarios
invoked to explain the MDR, such as harassment, which does change
galaxy structure, will have a different effects on the axis ratio
distribution of quiescent galaxies in high-mass halos.

Finally, we discuss our findings in the context of previous work and
identify slow gas stripping as the process that is likely responsible
for shaping the MDR (Section \ref{disc}).  This claim is then
explicitly shown to be consistent with the different properties of
spiral and S0 galaxies, and the increased fraction of elliptical
galaxies in dense environments.

\begin{figure}
\epsscale{1.2} 
\plotone{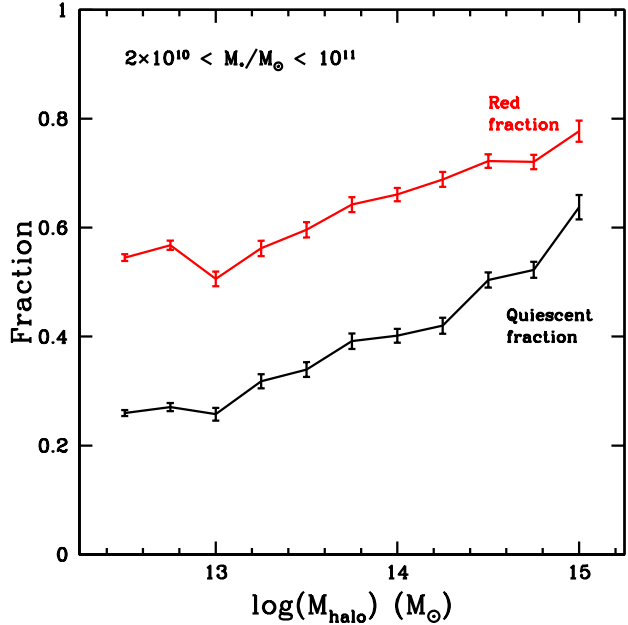}
\caption{Fraction of spectroscopically selected, quiescent
  (\textit{black lines and error bars}) and red (\textit{red lines and
    error bars}) galaxies with mass $2.5\times 10^{10} < M_*/\msol <
  10^{11}$ as a function dark matter halo mass from the \citet{yang07}
  group catalog.  The error bars indicate standard binomial errors in
  the fraction (these are henceforth used in all figures which show
  fractions).  Both the quiescent and red fraction increase with halo
  mass, and it is apparent that 'quiescence' is a more stringent
  criterion than 'redness'.}
\label{mdr}
\end{figure}

\section{Sample}\label{sample}
We select a sample of quiescent galaxies from Data Release 6 of the
SDSS \citep{adelman08}.  Our sample includes galaxies at redshifts
$0.04<z<0.08$ without detectable $[\rm{OII}]$ and $\rm{H}\alpha$
emission lines. The selection criteria are described and motivated in
full by \citet{graves09b}; but as opposed to that work, we do not
exclude galaxies with a low concentration index and galaxies that are
fit better by an exponential profile than by a \citet{devaucouleurs48}
profile, because this may exclude quiescent, yet disk-like galaxies.
As a consequence, our sample may include galaxies with star formation
in an extended disk outside the SDSS spectroscopic fiber which has a 3
arcsec diameter.  This, however, does not compromise our analysis as
the typical galaxy has a size that is similar to the spectroscopic
aperture.  Less than 20\% of the galaxies in our sample have sizes
that are more than twice this aperture.  The exclusion of all galaxies
with emission lines also excludes quiescent galaxies with active
nuclei.  Their number, however, is small, and make up a small fraction
of the population \citep[e.g.,][]{pasquali09a} that is negligible for
our purposes.

The axis ratios were obtained as described by \citet{vanderwel08c}.
Briefly, GALFIT \citep{peng02} is used to determine from the $r$ band
the radii, axis ratios, position angles, and total magnitudes,
assuming a \citet{devaucouleurs48} surface brightness profile.  The
smearing effect of the point-spread function is taken into account in
the model. Most galaxies have sizes comparable to the point-spread
function, that is, the global axis ratio is reliably recovered.  While
we use the de Vaucouleurs-derived values, we note that adopting
surface brightness models with a free \citet{sersic68} index does not
lead to a significantly different $b/a$ distribution. The systematic
difference between the two values is only 0.004 in the median, the
scatter is 0.075, and for less than 1\% of all galaxies the values
differ by more than 0.20.

The drawback of using the best-fitting axis ratio as a proxy for
morphology is that its value may not be very meaningful in case a
galaxy is a system with three components (say, bulge, disk, and bar)
which all contribute significantly to the light.  More detailed
modeling of the surface brightness profiles may alleviate this
problem.  However, this would be an extensive study in itself and,
moreover, prone to other, potentially more serious systematic errors
due to the limited information in the SDSS images of the galaxies in
our sample.

Stellar masses are estimated with a simple conversion from color to
mass-to-light ratio: $M/L_r=1.097\times (g-r) - 0.306 - 0.1$ taken
from \citet{bell03}, where $L_r$ and $g-r$ are computed at $z=0$, and
with a 0.1 dex shift downward to normalize all stellar masses to the
\citet{kroupa01} stellar initial mass function.  The assumed cosmology
is $(\Omega_{\rm{M}},~\Omega_{\Lambda},~h) = (0.3,~0.7,~0.7)$.

We match this sample with the galaxy group catalog constructed by
\citet{yang07}. Their method first uses a friends-of-friends algorithm
to identify the centers of potential groups, whose characteristic
luminosity is then estimated.  Using an iterative approach, the
adaptive group finder then uses the average mass-to-light ratios of
groups, obtained from the previous iteration, to assign a tentative
mass to each group.  This mass is then used to estimate the mass of
the underlying halo that hosts the group, which is in turn used to
determine group membership in redshift space.  Finally, each
individual group is assigned a halo mass, estimated from the group's
stellar mass, by ranking those stellar masses and the halo masses from
a numerical simulation of cosmological structure growth.  For more
details, we refer the reader to \citet{yang07}.

The group finder is optimized to maximize completeness while
minimizing contamination by interlopers.  The most massive galaxy of
each group is, by definition, the `central galaxy', and is usually
located near the geometric center of the group.  The other galaxies in
the group are 'satellite' galaxies.  However, the distinction between
'central' and 'satellite' galaxies is not of great importance for the
our analysis.

As \citet{yang07} constructed the catalog from SDSS Data Release 4
\citep{adelman06}, we do not have complete information for all
galaxies in our initial sample, which is selected from Data Release 6.
After cross matching, we have a sample of $\sim12,800$ quiescent
galaxies that are more massive than $M_*=2.5\times 10^{10}\msol$ (note
that dwarf galaxies are not considered here), and which also have been
assigned membership of a group as either a satellite or a central
galaxy.  This sample is used in this paper.  A complementary sample of
$\sim31,000$ star forming galaxies (i.e., those galaxies that do not
satisfy the selection criteria for quiescence given above, and in the
same redshift range, with known colors, stellar masses, and group
masses) is used in this paper to quantify the fraction of quiescent
galaxies in different environments, that is, in halos with different
masses.

\section{A Physical Description of the Morphology-Density
  Relation}\label{results}
\subsection{Dark Matter Halo Mass and the Quiescent Galaxy
  Fraction}\label{secmdr}
In Figure \ref{mdr}, we show the fraction of red\footnote{Our
  definition of a red galaxy is a galaxy which is less than 2 standard
  deviations bluer than the center of the red sequence, i.e., galaxies
  that satisfy $g-r > 0.07\times \log(M_*/10^{11}\msol)+0.75$, where
  $g-r$ is computed at $z=0$, are red.} and quiescent galaxies as a
function of halo mass.  Essentially all quiescent galaxies are red,
but the reverse is not the case.  Late-type galaxies with less than
average star formation activity may satisfy our color criterion, and
edge-on spiral galaxies may be relatively red because of extinction.
Moreover, galaxies with genuinely low star formation rates but with
some level of nuclear activity will be included in the red sample, but
not in the quiescent sample.  Generally speaking, it is important to
keep in mind that a sample of galaxies selected by optical color
consists of galaxies with a wide range of properties and should not be
equated with a sample of quiescent galaxies.  We focus on quiescent
galaxies for the pragmatic reason that it is well defined to address
the question at hand: are environmental trends due to the loss of cold
gas and the consequential decline in star formation activity?

Despite these complications, for both the red and the quiescent
sample, the observed trend resembles the picture that was originally
sketched by the MDR, that is, the fraction of red/quiescent galaxies
increases with halo mass.  However, no distinction between high- and
low-mass galaxies is being made here.  As a consequence, the observed
trend includes, in addition to the actual relationship between galaxy
properties and environment, if it exists, the relation between
color/star formation activity and galaxy mass, and between galaxy mass
and environment.

\begin{figure}
\epsscale{1.2} 
\plotone{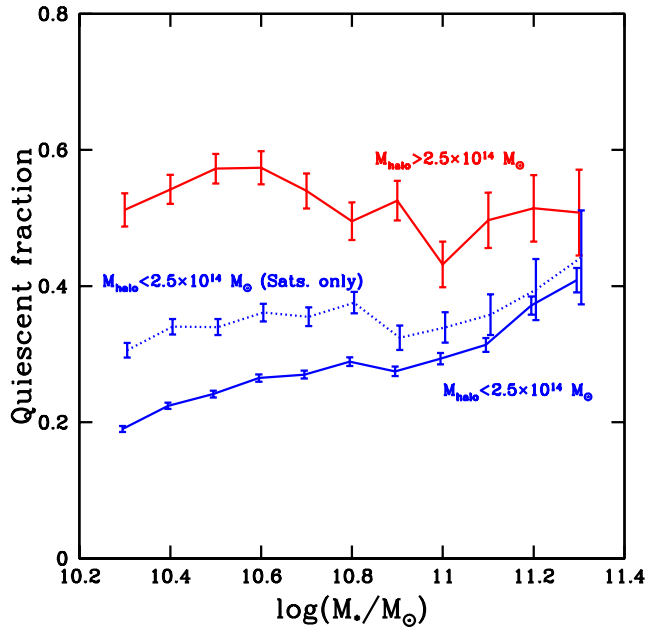}
\caption{Fraction of quiescent galaxies as a function of galaxy
  stellar mass. The red line indicates galaxies in high-mass halos
  (these are essentially all satellite galaxies); the solid blue lines
  indicate galaxies in low-mass halos; the dotted blue line shows
  satellite galaxies in low-mass halos.  The quiescent galaxy fraction
  is significantly higher in high-mass halos than in low-mass halos,
  even at fixed galaxy stellar mass.  This is partially, but not
  mostly, due to the difference between central and satellite
  galaxies: satellites in high-mass halos are different from
  satellites in low-mass halos.}
\label{frac}
\end{figure}

To gain a clearer perspective, we split the sample into galaxies in
high-mass halos ($M_{\rm{halo}}>2.5\times 10^{14}~\msol$) and galaxies
in low-mass halos ($M_{\rm{halo}} < 2.5\times 10^{14}~\msol$).  Our
choice for this value, which corresponds to the mass of a relatively
low-mass cluster with a velocity dispersion of $\sim 400-500~\kms$, to
distinguish between high- and low-mass halos is motivated in
\S\ref{disks}.  In Figure \ref{frac} we compare the fractions of
quiescent galaxies at fixed galaxy mass in high- and low-mass halos
(Figure \ref{frac}).  Now that we have taken out the galaxy mass
dependence, and, to first order, the halo mass dependence, the
intrinsic environmental dependence of star formation activity is
revealed.  The same trend, but then for color instead of
star formation activity, was found before by \citet{weinmann06a},
justifying the not entirely obvious assumption that optical color can
be used to directly trace star formation activity.  The trend seen in
Figure \ref{frac} is partially driven by the different properties of
central and satellite galaxies \citep[e.g.,][]{vandenbosch08a,
  pasquali09a, skibba09b}, but this is not the dominant factor. Also
for satellite galaxies we see that those in high-mass halos are more
frequently quiescent than in low-mass halos.  This can be seen in
Figure \ref{frac}, where the fraction of quiescent satellites is
smaller in low-mass halos than in high-mass halos. We note that
essentially all galaxies in high-mass halos are satellites.



\begin{figure*}
\epsscale{1.1} 
\plottwo{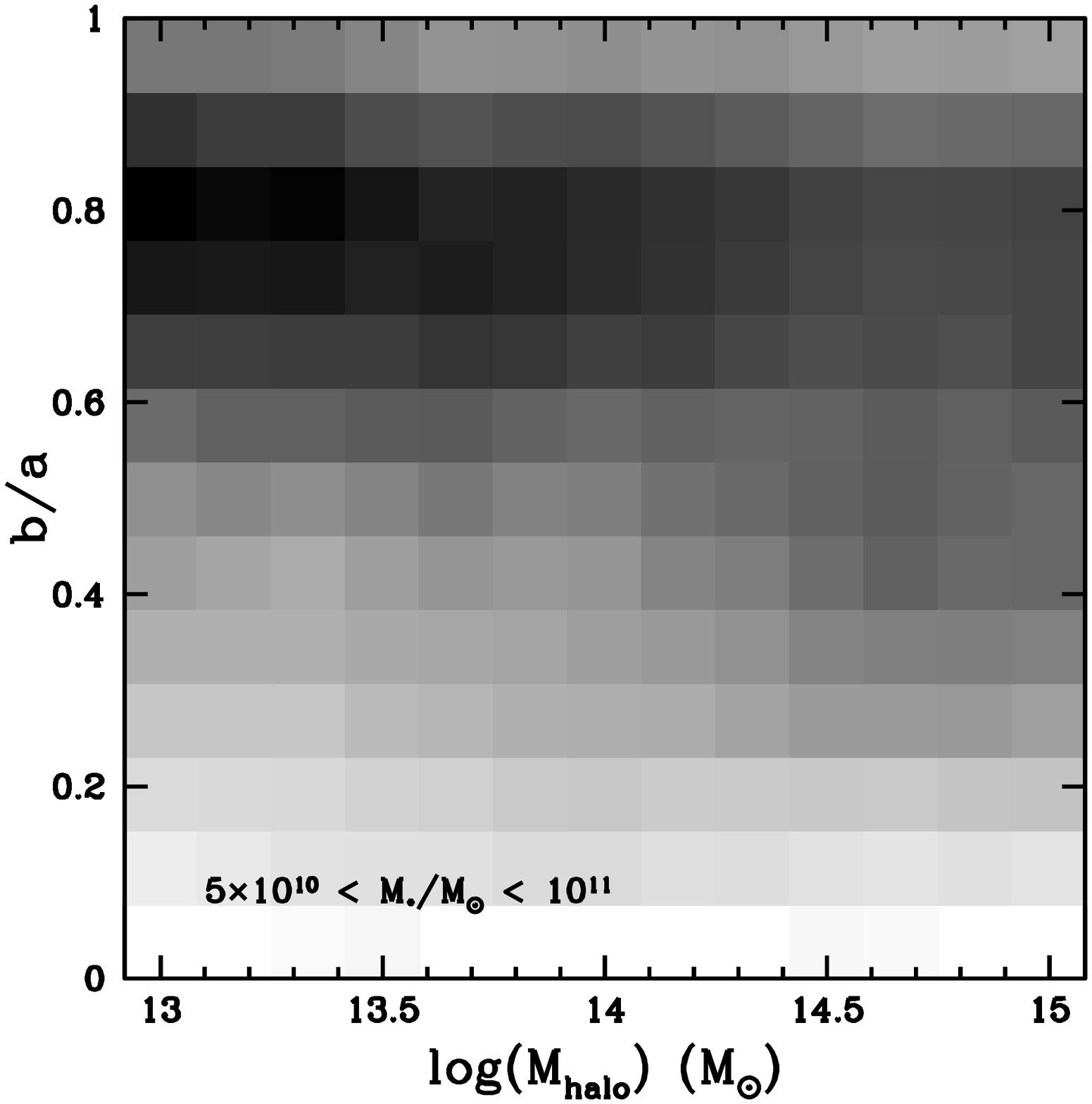}{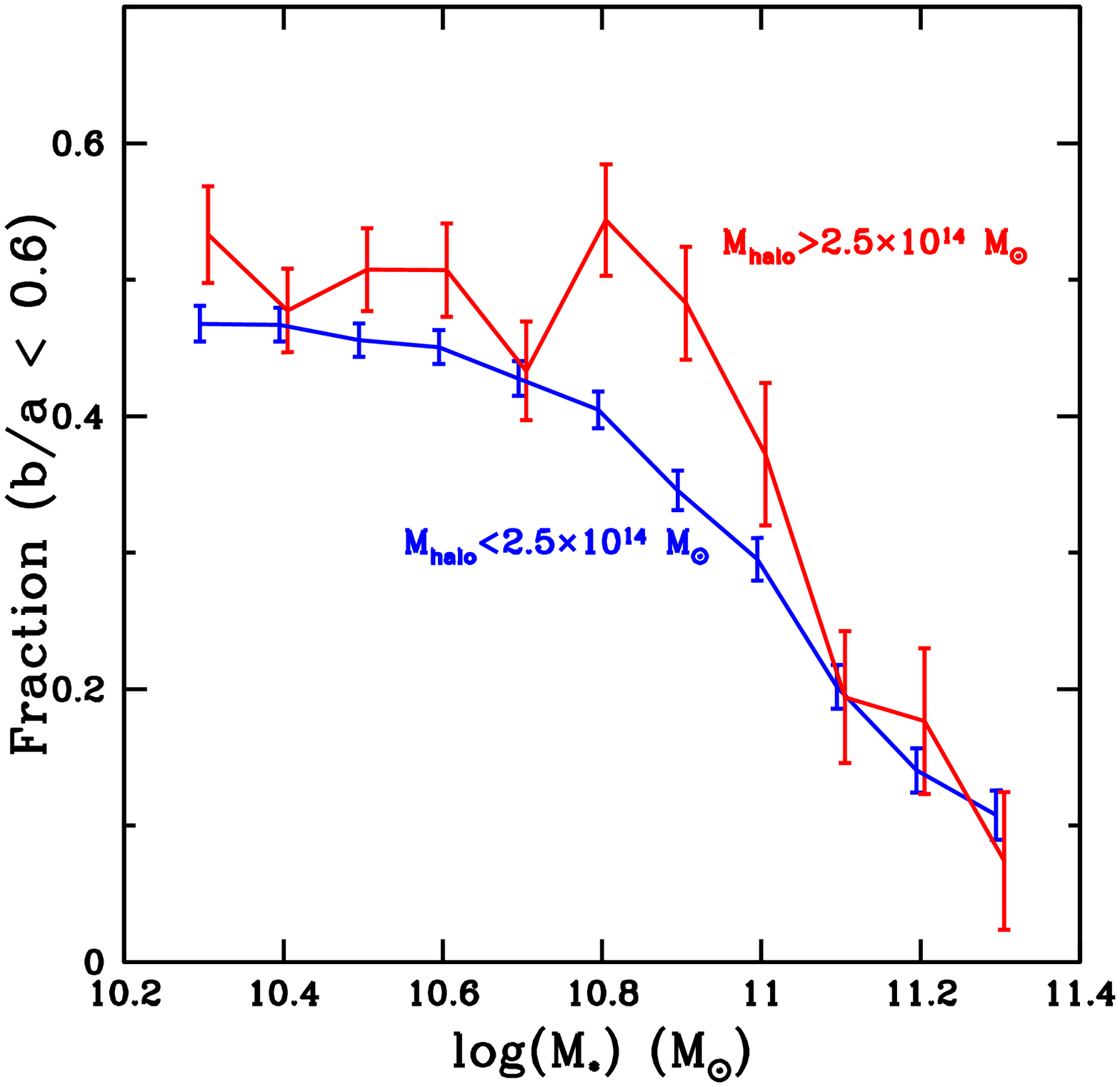}
\caption{\textit{Left:} axis ratio ($b/a$) distribution as a function
  of dark matter host halo mass for quiescent galaxies in the mass
  range $5\times 10^{10} < M_*/\msol < 10^{11}$. The gray scale
  indicates the frequency of $b/a$, where we normalize to unity the
  total number of galaxies in each stellar mass bin, the width of
  which is indicated by the grid size.  Smoothing in both directions
  is applied to reduce the effects of the shot noise.  The presence of
  the dark region near the top left corner implies that quiescent
  galaxies in low-mass halos are typically round. The $b/a$
  distribution at high halo masses is more uniform, with a higher
  fraction of elongated, that is, disky, quiescent galaxies.
  \textit{Right:} fraction of quiescent galaxies with axis ratios
  $b/a<0.6$ as a function of galaxy mass.  The red line represents
  galaxies in massive halos; the blue line represents galaxies in
  low-mass halos.  Error bars represent the Poisson uncertainty.
  Quiescent galaxies with $M_*\lesssim 10^{11}~\msol$ in high-mass
  halos are significantly more often elongated, that is,
  disk-dominated, than their counterparts in low-mass halos.
}
\label{frachalo}
\end{figure*}

\subsection{High-mass halos and quiescent galaxies with prominent
  disks}\label{disks}
Now that we have established that galaxies in high-mass halos are more
often quiescent than galaxies in low-mass halos, we address the
question what type of galaxy drives this trend.  We use our axis ratio
($b/a$) measurements as a tracer of the frequency of disks in the
population of quiescent galaxies.  Hence, we can test the hypothesis
that the truncation of star formation in high-mass halos is associated
with the destruction, preservation, or production of thin, stellar
disks.  Axis ratios do not provide a very sensitive test, because of
projection effects, but they are very robust, with negligible
systematic and random measurement errors (see Section \ref{sample}).

In Figure \ref{frachalo} (left), we show the $b/a$ distribution for
quiescent galaxies in a narrow range of stellar mass, as a function of
halo mass.  The halo mass range shown here, $M_{\rm{halo}} = 10^{13} -
10^{15}\msol$, probes a large range of environments, from groups with
a few $L^*$ galaxies to large clusters with hundreds of $L^*$ galaxies
and velocity dispersions of $\sim 1000~\kms$, such as the Coma
cluster.  At halo masses $M_{\rm{halo}} < 10^{13}\msol$ the group
catalog is essentially a catalog of relatively isolated galaxies, the
halos of which have been assigned masses that correspond directly to
the masses of those galaxies.  We omit this mass range from the figure
because it would not show trends with environment/halo mass, but
rather the relationship between galaxy mass and axis ratio.

Figure \ref{frachalo} reveals that high-mass halos host a larger
fraction of quiescent galaxies with small $b/a$ than low-mass halos,
which are dominated by relatively round systems.  In particular for
galaxies with masses ranging from $M_*=5\times 10^{10}\msol$ to
$M_*=10^{11}\msol$ the trend is highly significant (see Figure
\ref{frachalo}, right).  In this galaxy mass range, a transition in
the $b/a$ distribution occurs around $M_{\rm{halo}} = 2.5\times
10^{14}\msol$, the mass of a relatively small cluster, which explains
our choice to distinguish between high- and low-mass halos at this
particular value, first introduced in Section \ref{secmdr}, and used
throughout this paper.

Very few high-mass galaxies have pronounced disks, as shown recently
by \citet{vanderwel09b}.  Massive galaxies in high-mass halos do not
form an exception to this general rule, and, as a result, the axis
ratio distribution is not seen to vary with halo mass.

Because the majority of galaxies with relatively small masses
($M_*<5\times 10^{10}\msol$) have disks, it is harder to distinguish
an excess population of disks in high-mass halos.  A small difference,
however, is still apparent, as can be seen in Figure \ref{frachalo}
(right).  The relatively small effect on the axis ratio distribution
does not imply that environmental processes are weaker at these lower
masses; on the contrary, environmental processes are more pronounced
at lower masses, as can be seen in Figure \ref{frac}.

\begin{figure*}
\epsscale{1.2} 
\plotone{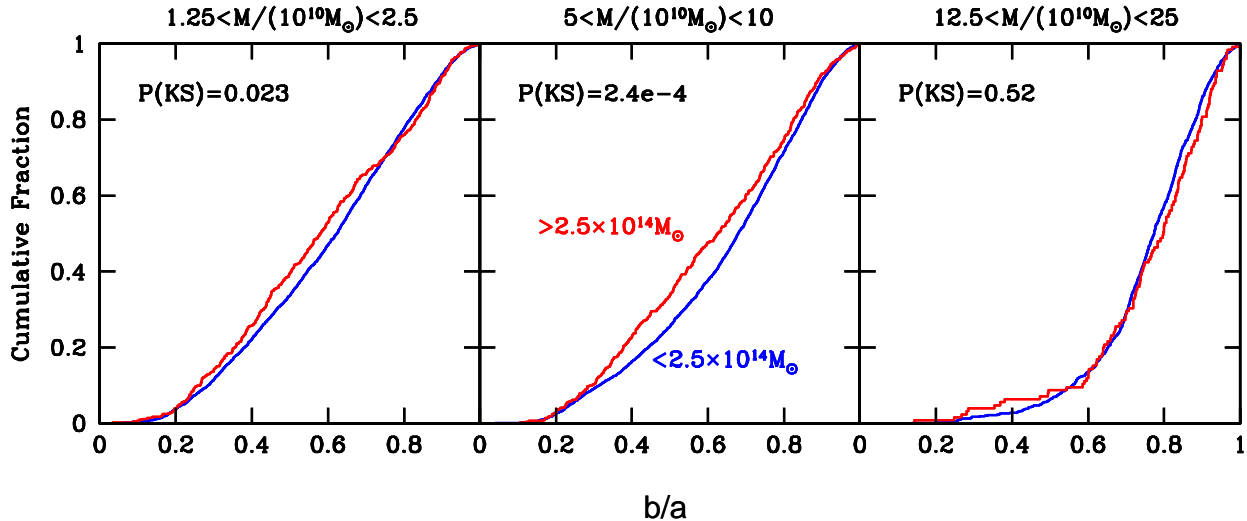}
\caption{Cumulative axis ratio distribution for quiescent galaxies in
  three different galaxy mass bins. The red lines reflect the $b/a$
  distribution of galaxies in high-mass halos; the blue lines indicate
  the $b/a$ distribution of galaxies in low-mass halos, as labeled in
  the middle panel. $P(KS)$ is the probability that the $b/a$ values
  of galaxies in low- and high-mass halos are drawn from the same
  distribution, using the standard Kolmogorov-Smirnov test.  For
  high-mass galaxies (\textit{right}) the two $b/a$ distributions are
  statistically indistinguishable.  For galaxies with masses $M_* <
  10^{11}\msol$, the probability that the $b/a$ distributions in high-
  and low-mass halos are the same is very small: high-mass halos
  contain an excess population of disk-dominated quiescent galaxies
  compared to low-mass halos.}
\label{cumu}
\end{figure*}

In order to formalize the trends seen in Figure \ref{frachalo}, we
analyze the cumulative $b/a$ distributions for different ranges in
galaxy mass.  These are shown in Figure \ref{cumu}.  As was already
indicated above, the $b/a$ distributions of massive galaxies in high-
and low-mass halos are statistically indistinguishable.  For galaxies
with masses below $M_*=10^{11}\msol$, however, there is a notable and
significant difference.  According to the standard Kolmogorov-Smirnov
test it is highly unlikely that quiescent galaxies have the same $b/a$
distributions in low- and high-mass halos, in the sense that high-mass
halos more frequently host disky quiescent galaxies (see Figure
\ref{cumu}).

Recently, \citet{vandenbergh09} attempted to identify such a
correlation between flatness and environment, but did not find a
statistically significant trend.  A sample of several hundred galaxies
is perhaps insufficient to disentangle the large variety of underlying
correlations between global properties of galaxies.  The sample that
we use in this paper is almost 2 orders of magnitude larger, which
adequately remedies this problem.

\subsection{Disky quiescent galaxies drive the morphology-density
  relation}
\label{mdrdisks}
In Section \ref{secmdr}, we established that high-mass halos have a
higher fraction of quiescent galaxies - at a given galaxy mass - than
low-mass halos (see Figure \ref{frac}).  In Section \ref{disks} we
found that quiescent galaxies in high-mass halos are more frequently
disky than in low-mass halos (see Figures \ref{frachalo} and
\ref{cumu}).  These two observations could be physically related, in
which case the increased quiescent fraction in high-mass halos is
directly caused by an increased fraction of quiescent disk galaxies;
see also the discussion by \citet{vandenbosch08a}.

We test this hypothesis with a simple model. The idea is that the
quiescent population in high-mass halos consists of galaxies that
would be quiescent even if they had been located in less massive
halos, that is, regardless of their environment, and an additional
population that is quiescent as a direct consequence of their
environment.  We assume that the latter population is the result of
gas stripping, which leaves stellar disks intact and does not alter
the structural properties of galaxies.  We discuss this assumption
below.

The expected $b/a$ distribution of quiescent galaxies in high-mass
halos is inferred by adding these two populations: (1) a population
with the $b/a$ distribution of similarly massive quiescent galaxies in
low-mass halos, which represent galaxies that are quiescent regardless
of halo mass
and (2) a population of galaxies, projected along random
lines-of-sight, with the same intrinsic axis ratios as similarly
massive spiral galaxies (i.e., non-quiescent $L^*$ galaxies), which
represent galaxies that are quiescent because of their environment.
According to \citet{padilla08}, $L^*$ spiral galaxies have intrinsic
axis ratios that are normally distributed, with a mean of 0.26 and a
scatter of 0.06. The motivation for this choice is that such spiral
galaxies will, upon entering a massive halo, eventually cease to form
stars without changes in structure.

Besides the $b/a$ distributions of the two populations in the model,
we also have to choose the relative numbers of galaxies in the two
populations.  We choose these relative numbers such that we precisely
match the increase in the quiescent galaxy fraction from low- to
high-mass halos, which we described in Section \ref{secmdr} and showed
in Figure \ref{frac}.  Hence, if our gas-stripping hypothesis is
correct, then the model $b/a$ distribution is consistent with the
observed $b/a$ distribution in high-mass halos.

\begin{figure*}
\epsscale{1.1} 
\plottwo{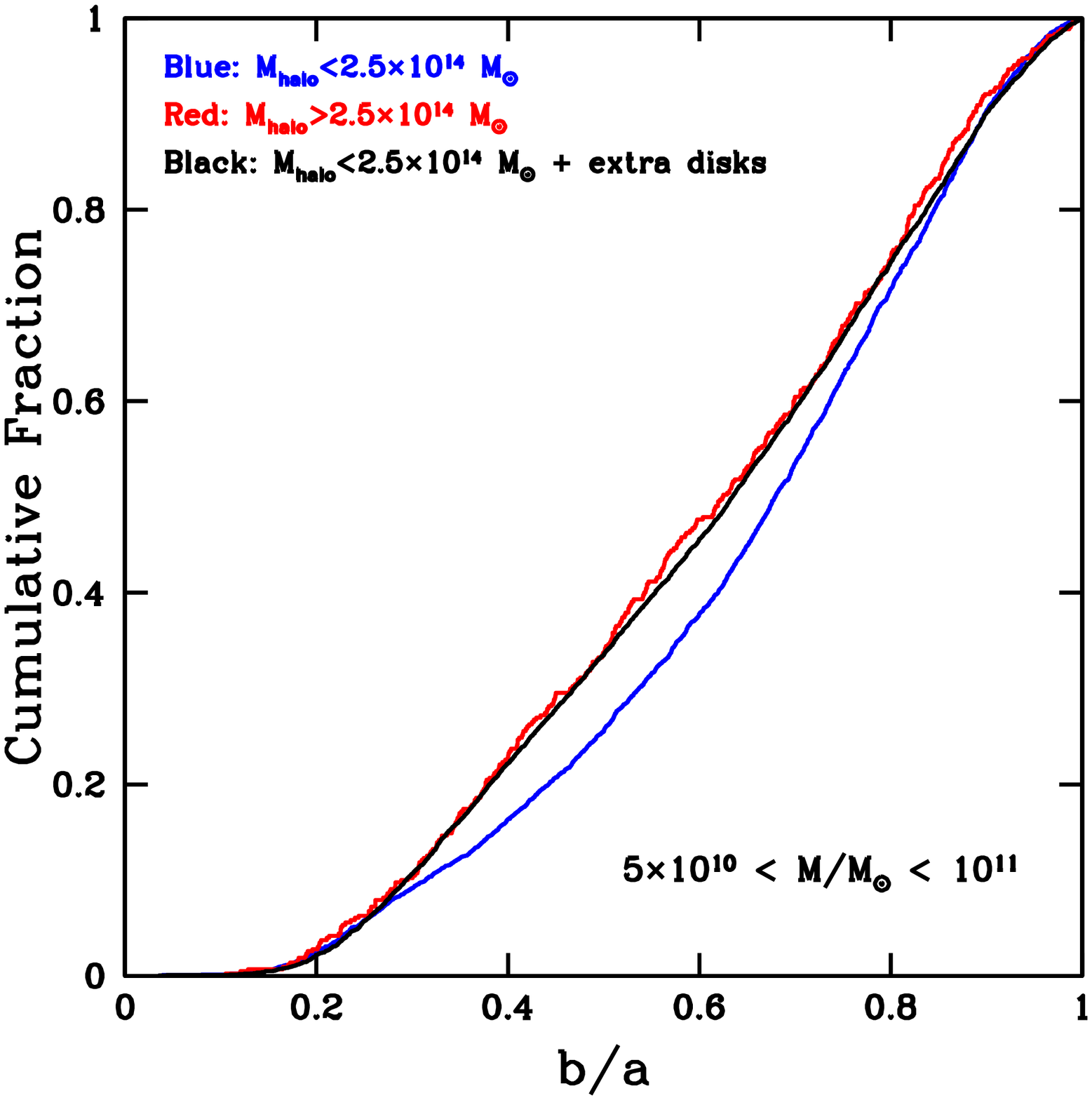}{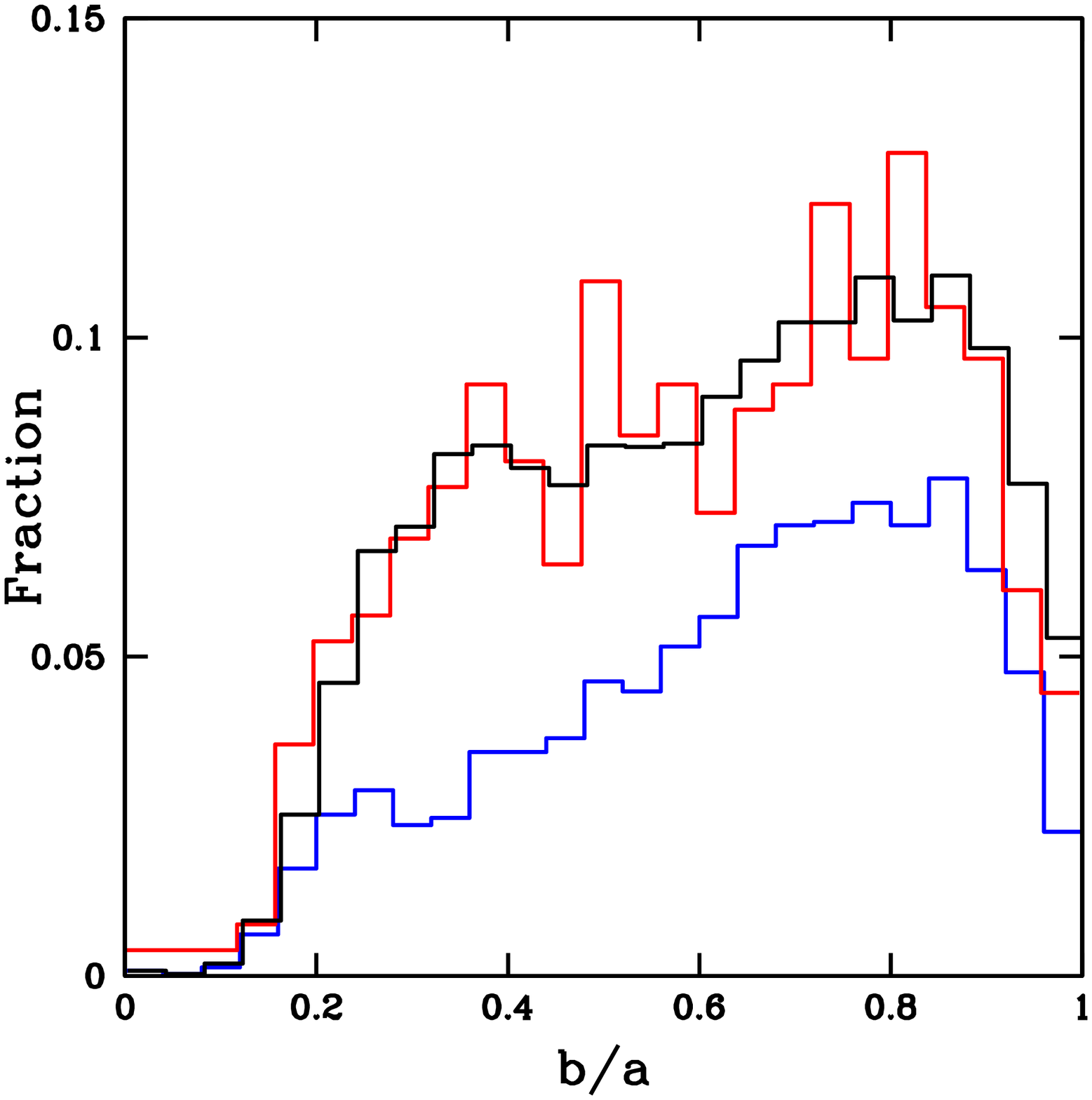}
\caption{Cumulative distribution (\textit{left}) and histogram
  (\textit{right}) of axis ratios of quiescent galaxies with masses in
  the range $5\times10^{10}<M_*/\msol<10^{11}$.  The solid red lines
  reflect the $b/a$ distribution of galaxies in high-mass halos; the
  blue lines reflect the $b/a$ distribution of galaxies in low-mass
  halos.  The black line is the $b/a$ distribution in low-mass halos
  (i.e., the blue line) augmented by a population of 'extra' quiescent
  disk galaxies.  The latter is generated by projecting along random
  lines of sight a population of disk galaxies with the same intrinsic
  axis ratio distribution as $L^*$ spiral galaxies (see the text for
  details).  The relative number of these extra disk galaxies is
  chosen such that their addition makes up for the difference in the
  quiescent galaxy fractions in high- and low-mass halos (see Figure
  \ref{frac}) for the galaxy stellar mass range considered here.
  These extra disk galaxies can be thought of as galaxies whose star
  formation has been truncated as a results of some environmental
  process that leaves the stellar disk intact.  The resulting model
  $b/a$ distribution is statistically indistinguishable from the
  observed $b/a$ distribution in high-mass halos: the
  Kolmogorov-Smirnov test yields a probability of 0.83 that the
  distributions are the same.  Therefore, the increased fraction of
  quiescent galaxies in high-mass halos seen in Figure \ref{frac} is
  likely caused by the increased fraction of disky, quiescent galaxies
  seen in Figures \ref{frachalo} and \ref{cumu}.}
\label{extr}
\end{figure*}

In Figure \ref{extr}, we compare the predicted and observed $b/a$
distributions in high-mass halos for galaxies in the mass range
$5\times10^{10} < M_*/\msol < 10^{11}$.  The population of 'extra',
disky quiescent galaxies added to the $b/a$ distribution of galaxies
in low-mass halos changes the $b/a$ distribution in such a manner that
the $b/a$ distribution in high-mass halos is accurately reproduced.
There is no significant difference between the predicted and observed
$b/a$ distributions in high-mass halos: according to the
Kolmogorov-Smirnov test, the probability that the distributions are
the same is $0.83\pm 0.06$.\footnote{In a Monte Carlo simulation, this
  is the average of 1000 realizations of the random inclination
  distribution of the 'extra' disky, quiescent galaxy population. This
  produces a scatter of 0.06 in the inferred probability that the
  distributions are the same.}  This is to be compared with the low
probability ($2.3\times 10^{-4}$) that the $b/a$ distributions in low-
and high-mass halos are the same, a remarkable success for such a
simple model.

We repeat the exercise for galaxies with lower masses (the left-hand
panel in Figure \ref{cumu}).  The difference between low- and
high-mass halos is less pronounced and less significant here.  Yet,
adding an extra population of faded spirals to the $b/a$ distribution
observed in low-mass halos, as described above, we again reproduce
with good confidence the observed $b/a$ distribution in high-mass
halos.

This model, however, does not work for high-mass galaxies
($M>10^{11}\msol$).  In the right-hand panel of Figure \ref{cumu} can
be seen that the $b/a$ distributions of massive galaxies are not
different in low- and high-mass halos, yet, Figure \ref{frac} shows
that even for such massive galaxies there is a small, but significant
difference in the quiescent fraction.  As a consequence, modeling the
additional quiescent galaxies in high-mass halos with a population of
faded spiral galaxies does not reproduce the observed $b/a$
distribution in high-mass halos.  We discuss this further in Section
\ref{disc}.

We note that none of the above results change if we restrict the
analysis to satellite galaxies alone.  The increased fraction of
quiescent satellite galaxies in high-mass halos (see Figure
\ref{frac}) is well described by an increased number of quiescent,
disk-dominated galaxies, analogous to our above description of the
entire population of quiescent galaxies.

The assumption in all of the above is that the axis ratio does not
change when a star forming spiral galaxy transforms into a quiescent,
disky galaxy.  More explicitly, we assume that the stellar mass
distribution is not significantly affected by the process that
truncates the star formation.  However, it is good to keep in mind
that tidal interactions may thicken the disk, and radial mixing, on
the long term, can change the bulge-to-disk ratio \citep[see][for an
overview]{boselli06}. We also neglect changes in surface brightness
profiles due to changes in mass-to-light ratio gradients caused by
either stellar population effects or rearrangement of dust.  However,
the fact that we can model the axis ratio distribution in high-mass
halos by assuming that the structure of a galaxy does not change when
its star formation activity is truncated, suggests that our assumption
is reasonable.

This section presents the central result of this study.  In summary,
we connect the increased fraction of quiescent galaxies in high-mass
halos (see Figure \ref{frac}) to the increased fraction of
disk-dominated, quiescent galaxies in high-mass halos (see Figures
\ref{frachalo} and \ref{cumu}).  We demonstrate quantitatively that
the increased fraction of quiescent galaxies in high-mass halos is in
fact driven by the population of disk-dominated galaxies (Figure
\ref{extr}).  In the following section, we discuss the ramifications
of these findings in the context of the historical discussion of the
MDR and in light of other recent results.

\section{Discussion}\label{disc}

\subsection{The definition of 'morphology' and its interpretation}

The traditional morphological classification of galaxies is based on
two observable properties: smoothness of the light distribution and
the presence/absence of a substantial disk \citep[e.g.,][]{sandage61}.
Physically speaking, smoothness reflects star formation activity,
whereas the presence of a disk implies a dynamically cold component.
Because these properties are correlated with each other, and with a
host of other galaxy parameters, morphology is not a fundamental
galaxy property, which renders morphological studies difficult to
interpret.  We refer to \citet{blanton09} for a recent, general
overview of the field.

In any case, these two observables, smoothness and diskiness, divide
the general population into three broad classes: ellipticals (smooth,
negligible disks), lenticulars or S0s (smooth, significant disks), and
spirals (not smooth, significant disks).\footnote{We are only
  considering galaxies with masses more than 20\% of that of an $L^*$
  star galaxy, that is, the following discussion does not apply to
  dwarf galaxies, which constitute a somewhat separate class of
  objects.}  Smoothness is related to the lack of star formation
activity; hence, our selection of quiescent galaxies is equivalent to
selecting ellipticals and S0s, that is, early-type galaxies.

From the early SDSS studies by, for example, \citet{kauffmann03b} and
\citet{hogg04}, we learned that diskiness is primarily related to
galaxy mass, and also that its correlation with environment is weak
\citep{kauffmann04, blanton05b}.  Smoothness, on the other hand, has
been demonstrated to correlate with environment as well
\citep{vanderwel08a}, which should be interpreted as an environmental
dependence of star formation activity \citep[see also][who all find a
correlation between color/star formation and environment]{lewis02,
  gomez03, balogh04, tanaka04, weinmann06a, baldry06, park07}.  Hence,
we have that the two physical properties that make up morphology
behave differently as a function of fundamental galaxy properties,
reconciling the result that whereas morphology is sensitive to
environmental factors, structure is not, or at least less so
\citep{vanderwel08a, bamford09, skibba09c}.


\subsection{Gas stripping and the morphology-density relation}

The degree to which the star formation activity of a galaxy is
environmentally affected depends on halo mass \citep{weinmann09a}.
Satellite galaxies in high-mass halos have different color profiles
and are generally redder than their equally massive counterparts in
low-mass halos.  Model constraints on the amount of time that galaxies
have spent as satellites and the observed correlation between halo
mass and satellite color imply that the decrease in star formation
activity is gradual, taking place on a time scale of several Gyr.
This is reinforced by the dependence of star formation activity of
satellite galaxies in massive halos on cluster-centric distance
\citep{vonderlinden09}, and, even more generally, by the non-zero
fraction of blue satellite galaxies \citep{kang08, font08, skibba09a}.

These results are inconsistent with a scenario in which the entire
interstellar medium is rapidly stripped as a result of ram pressure.
Increasingly sophisticated simulations indicate that the time scale
for gas loss through ram-pressure stripping is longer than expected on
the basis of simple, analytical estimates \citep[e.g.,][]{roediger07}.
The picture that is emerging is that star formation continues after
infall, but gradually declines as the fuel for star formation is
removed relatively slowly and gently by the stripping of hot gas
outside the cold disk or atomic hydrogen in the outer parts
\citep[e.g.,][]{koopmann04a}.  However, we note that the physical
process at work here is still ram-pressure stripping, but that this
only efficiently acts on part of the interstellar medium.

Explicit evidence for this process is provided by the observation of
an excess of red, late-type galaxies in the outskirts of a massive
cluster \citep{wolf09, gallazzi09}, which have lower star formation
rates than late-type galaxies in the general field.  More generally,
the clustering properties of red spiral galaxies imply that such
galaxies are often satellite galaxies in massive halos
\citep{skibba09c}.  Interestingly, such 'anemic' spirals were
identified and discussed early on, by \citet{vandenbergh76} and have
been suggested as an explanation for the MDR \citep{koopmann98}.

The end-result of the gas-stripping process, that is, the descendant
of a red spiral galaxy with a declining star formation rate, is a
quiescent, disk-dominated galaxy of which the bulk of the stellar
population is at least several Gyr old, in agreement with
spectroscopically inferred age estimates of quiescent galaxies in
general \citep[e.g.,][]{gallazzi05, graves09a}.  These conclusions,
together with the dependence on halo mass of the average amount of
time that a galaxy has spent as a satellite, may explain the
correlation between the stellar population ages of red galaxies and
environment \citep{cooper09}.

Our results provide strong evidence for the presence of such
``starved'', disky galaxies in massive halos.
Moreover, our results imply that these galaxies represent the
increased fraction of quiescent galaxies in high-mass halos.  Hence,
our results, combined with those discussed above, imply that a gradual
removal of the gaseous interstellar medium of spiral galaxies is
responsible for the MDR.  This conclusion is physically motivated by
the observed decreased star formation in satellite galaxies and the
time scale associated with this phenomenon
\citep[e.g.,][]{weinmann09a}, and quantitatively matched by the
observed increase in the fraction of disk-dominated, quiescent
galaxies in massive halos (this paper).  

\subsection{Are S0 galaxies stripped spiral galaxies?}

In this paper we argue that slow gas stripping through interaction
with the intergalactic medium causes spiral galaxies to stop forming
stars and eventually turn in disky, quiescent objects, that is, S0
galaxies.  Furthermore, we claim that this explains the existence of
the MDR for galaxies with masses $M>2.5\times 10^{10}\msol$.  Note
that in this picture, galaxies do not undergo structural changes, in
the sense that their bulge-to-disk ratios stay the same.

We stress that we do not claim that all S0 galaxies formed through
this process.  The existence of S0 galaxies and existence the MDR do
not necessarily have the same explanation, which was already noted by
\citet{postman84}. Only $\sim15\%$ of the galaxies in our sample are
located in dark matter halos with masses $M_{\rm{halo}} > 2.5\times
10^{14}\msol$, where the increased presence of quiescent disk galaxies
is noticeable.  For this sub-population, the fraction of quiescent
galaxies is 1.5-2 times higher than in lower-mass halos (see Figure
\ref{frac}), implying that only 5\%-10\% of all quiescent galaxies are
quiescent as a result of differential environmental effects, by which
we mean those environmental effects that do not or less efficiently
act in lower-mass halos.  Although these are rough estimates, it is
safe to conclude that the process that is responsible for the MDR is
not the dominant contributor to the truncation of star formation and
the formation of the red sequence in general.  

It is not surprising, then, that the MDR is as weak as it is:
morphological fractions change only by a factor 2 over many orders of
magnitude in environmental density, as was pointed out before, by,
e.g., \citet{dressler04}.  The star formation and structural
properties of galaxies depend much more strongly on internal galaxy
properties (mass, velocity dispersion, and surface mass density) than
on their environment.  Many central galaxies are quiescent and many
quiescent satellite galaxies had their star formation truncated before
their becoming satellites \citep{vandenbosch08a}.

Although low-mass satellites in low-mass halos could still be the
result of gas stripping, it seems unavoidable that other mechanisms
that truncate star formation are important.  Merging may explain the
cessation of star formation by means of gas exhaustion through
enhanced star formation activity during the merger phase.  Stellar
disks are not destroyed in the case of minor merging, although the
bulge-to-disk ratio can increase \citep[e.g.,][]{bekki98, naab99}, and
a sequence of minor mergers can result in an elliptical galaxy
\citep{bournaud07}.  It is generally accepted that major merging
results in round remnants \citep{toomre72, barnes96}, although such
events can also lead to the formation of flattened, rotating systems,
especially if the progenitors are relatively gas rich \citep{cox06}.


Whether merging will turn out to be the main mechanism to produce
quiescent galaxies remains to be seen.  However, quiescence seems to
be related to the presence of a bulge \citep[e.g.,][]{bell08},
produced by merging or otherwise.  In its barest form, this
relationship manifests itself through the correlation between
structure and star formation activity.  Hence, it is not at all
surprising that S0s, which are quiescent, have larger bulges than
star forming spirals.  In other words, S0s are not, generally
speaking, gas-stripped spirals.  However, as explained above, this is
not at odds with our claim that gas stripping results in the existence
of the MDR.


If merging is an important driver of bulge growth, then differences
between S0s and spirals are expected.  These differences include the
offset in the \citet{tully77} relation between spirals and S0s
(M. Williams et al., in prep), the prominence of thick disks in S0s
compared to spirals \citep[e.g.,][]{burstein79d}, and the lower bar
fraction for S0s \citep{aguerri09, laurikainen09}.  The accretion of
satellite galaxies/minor merging can destroy bars and cause thick
disks to grow more prominent, either through heating up the
pre-existing thin disk \citep[e.g.,][]{quinn93} or by depositing
tidally stripped debris from the accreted systems at large scale
heights \citep[e.g.,][]{gilmore02, martin04}.  In addition, minor
accretion events in the absence of gas is expected to lead to a
thicker disk than in the presence of gas \citep[e.g.,][]{moster09}.
Hence, even similar satellite accretion histories for S0 and spiral
galaxies can lead to differences between the thick disk components,
and the general prominence of the surviving disk.

In summary, the global differences between S0 and spiral galaxies do
not argue against slow gas stripping as an explanation for the MDR.
Most quiescent galaxies, including those with prominent disks, are
\textit{not} the result of differential environmental processes, that
is, processes that do not act (efficiently) in low-density
environments/low-mass halos.  If environmental processes are
important, they act efficiently in all environments.  However, it
seems inescapable to conclude that other truncation mechanisms, likely
associated with merging, are important, because bulge growth and the
truncation of star formation go hand in hand.

\subsection{Elliptical galaxies and environment}

Because the increase in the quiescent galaxy fraction with halo mass
can be fully explained by the increased fraction of intrinsically
flattened galaxies (Sec 3.2), it follows that the fraction of
intrinsically round quiescent galaxies (elliptical galaxies) does not
change with halo mass, at least, for galaxy masses below
$10^{11}\msol$.  At first sight, this seems to be at odds with the
increased fraction of elliptical galaxies, with respect to the total
population, at high local density \citep{dressler80}.

\citet{whitmore93} showed that galaxy morphology varies strongly with
position within a cluster.  They argue that the result from Dressler
can be explained by a higher fraction of elliptical galaxies in the
cluster core (within the central 0.25 Mpc) than elsewhere.  Using the
distance-to-group-center estimates from \citet{yang07}, we check
whether we can reproduce the trend shown by \citet{whitmore93}.  In
order to roughly match the properties of the galaxies in the sample
used by \citet{dressler80} and \citet{whitmore93}, we select galaxies
from our sample with mass $M_*>6\times 10^{10}~\msol$ in groups with
mass $M_{\rm{halo}}> 2.5\times 10^{14}~\msol$.  Quiescent galaxies in
that sample within the central 0.25 Mpc of the centers of their
respective groups are significantly rounder than galaxies at larger
distances from the group centers, which implies that elliptical
galaxies are more prevalent than S0s in cluster cores, and that our
sample shows the same trend as identified by \citet{whitmore93}.

It is important to note that the centers of groups and clusters tend
to be populated by the most massive galaxies.  This is relevant for
the present discussion because morphological type depends strongly on
galaxy mass: the axis ratio distribution implies that essentially all
high-mass galaxies are intrinsically round \citep{vanderwel09b}, that
is, those are all elliptical galaxies.  At lower masses, many galaxies
have significant disks (see also Figure \ref{cumu}).  This suggests
that the increased fraction of ellipticals in group centers includes a
contribution from two underlying trends: group centers host more
massive galaxies, which are, in turn, more often ellipticals.  We find
that this fully explains the dependence of the elliptical fraction on
distance to the group center: the axis ratio distribution of quiescent
galaxies with a given mass does not change with distance to group
center.

Recently, \citet{vonderlinden09} showed that satellite galaxies in
massive groups do not show a correlation between mass and distance to
the group center.  We find the same, and we only find mass segregation
if the central galaxies are included.  The implication is that the
\citet{whitmore93} result that elliptical galaxy fraction increases
toward the group center can be fully understood by distinguishing
between central and satellite galaxies.

In summary, the increased fraction of elliptical galaxies in dense
environments \citep{dressler80} and cluster cores \citep{whitmore93}
is not at odds with our claim that slow gas stripping of infalling
spiral galaxies explains the MDR at fixed galaxy mass.  The increased
elliptical galaxy fraction in galaxy cores is simply the consequence
of high-mass, elliptical galaxies preferring the inner regions.  We
show that, at fixed galaxy mass, S0 galaxies are more prevalent than
ellipticals in clusters compared to lower-density environments, an
issue that was not discussed in the early, seminal works by, e.g.,
\citet{dressler80} and \citet{postman84}.

\subsection{The morphological mix of high-redshift clusters of
  galaxies}
With the arrival of the \textit{Hubble Space Telescope}, it became
possible to study the morphologies of galaxies in distant clusters
\citep{dressler97, fasano00, treu03, postman05, smith05}, and
establish that the Hubble sequence and the MDR were already in place
at $z\sim 1$.  The same is the case for the underlying physical
correlations between star formation activity and environment
\citep[e.g.,][]{cooper07, patel09a, tran09}.

There are two reasons to suspect that the morphological mix may change
with look-back time. First, as cluster halos continue to assemble
hierarchically through the accretion of smaller halos, newly infalling
galaxies are added to the cluster population.  This infall and merging
process is directly observed at all redshifts, and must be an ongoing
process \citep[e.g.,][]{burns94, markevitch02}. This process adds
galaxies with field-like properties, which may differ from the already
present cluster galaxies.  Second, at higher redshift, the typical
star formation rate is higher for spiral galaxies \citep{lilly96,
  madau96}, which suggests that the morphological mix may change with
look-back time.

Indeed, the fraction of spiral/star forming galaxies is observed to
increase with redshift \citep[e.g.,][]{dressler97, smith05, postman05,
  poggianti06, simard09}.  In general, these studies suggest that the
population of S0 galaxies is rapidly built up over the last 8 Gyr from
infalling spiral systems \citep{dressler97, postman05, smith05}. It
was shown by \citet{holden07} and \citet{vanderwel07b} that this is
mainly due to the higher luminosity of spiral galaxies in distant
clusters compared to those in local clusters: in mass- selected
samples, little change is seen and the deficit of S0 systems appears
to be overestimated \citep{holden09a}.

The key result of \citet{vanderwel07b} is that, although there is
little evolution in the fraction of E+S0 galaxies at the stellar
masses of $L^*$ galaxies \citep[cf.][]{bundy09b}, there is still a
strong relation between morphology and the local environment
\citep[see also][]{tasca09}.  The persistence of this trend to higher
redshifts implies that galaxies that become satellites in more massive
halos transform from actively star forming to quiescent galaxies.
Such transformation are directly observed at intermediate redshifts
\citep{moran07}.

Our result shows that there is a population of galaxies in high-mass
halos that have a similar structure to star forming field spiral
galaxies, but lack the star formation.  \citet{vanderwel07b} show that
this population does not quickly appears at some given epoch.  From
this, we conclude that the assembly of the cluster population and the
evolution of its star forming properties are most naturally explained
by the gradual stripping of the interstellar medium.

\section{Summary and Conclusions}
We use stellar masses, star formation activity, axis ratios, and group
halo masses of galaxies in the SDSS to provide a physical
interpretation of the morphology-density relation and its origin.  Our
findings are as follows.
\begin{itemize}
\item{The fraction of galaxies with low specific star formation rates
    (quiescent galaxies) increases with halo mass. This also holds at
    fixed galaxy mass and for satellite galaxies (Section
    \ref{secmdr})}.
\item{Quiescent galaxies in high-mass dark matter halos
    ($M_{\rm{halo}} > 2.5\times 10^{14}\msol$) are significantly more
    often disk dominated than quiescent galaxies in lower-mass halos
    (Section \ref{disks}).}
\item{This additional population of disk-dominated quiescent galaxies
    quantitatively matches the increased fraction of quiescent
    galaxies in high-mass halos (Section \ref{mdrdisks}).}
\end{itemize}
Hence, our findings show that the morphology-density relation arises
as a result of the increased fraction of disk-dominated, quiescent
galaxies in high-mass halos, at the expense of disk-dominated, star
forming galaxies, which are more frequently found in low-mass halos.
Other studies (see Section 4.2) provide evidence that slow stripping
of the interstellar medium is the most likely explanation for the
decreased star formation activity of spiral galaxies in massive
groups.  We conclude that the slow stripping of gas from spiral
galaxies, which does not strongly alter its structural properties of a
galaxy, likely explains the MDR.

These findings are discussed in the context of the rich history of
studies on galaxy morphologies and their environmental dependence
(Section 4).  In particular, we demonstrate that our conclusions are
not incompatible with the suite of evidence that S0 galaxies and
spiral galaxies have systematically different properties.  Such
evidence has often, and correctly, been invoked to argue that S0
galaxies cannot, generally speaking, be stripped spiral galaxies.
Rather, the stripped, quiescent galaxies that drive the
morphology-density relation are only a small subset of the entire
population quiescent galaxies with disks.

\acknowledgements{We thank Frank van den Bosch and Anna Gallazzi for
  useful comments on the manuscript.

  Funding for the SDSS and SDSS-II has been provided by the Alfred
  P. Sloan Foundation, the Participating Institutions, the National
  Science Foundation, the U.S. Department of Energy, the National
  Aeronautics and Space Administration, the Japanese Monbukagakusho,
  the Max Planck Society, and the Higher Education Funding Council for
  England. The SDSS Web site is http://www.sdss.org/.

  The SDSS is managed by the Astrophysical Research Consortium for the
  Participating Institutions. The Participating Institutions are the
  American Museum of Natural History, Astrophysical Institute Potsdam,
  University of Basel, University of Cambridge, Case Western Reserve
  University, University of Chicago, Drexel University, Fermilab, the
  Institute for Advanced Study, the Japan Participation Group, Johns
  Hopkins University, the Joint Institute for Nuclear Astrophysics,
  the Kavli Institute for Particle Astrophysics and Cosmology, the
  Korean Scientist Group, the Chinese Academy of Sciences (LAMOST),
  Los Alamos National Laboratory, the Max-Planck-Institute for
  Astronomy (MPIA), the Max-Planck-Institute for Astrophysics (MPA),
  New Mexico State University, Ohio State University, University of
  Pittsburgh, University of Portsmouth, Princeton University, the
  United States Naval Observatory, and the University of Washington.
}

\bibliographystyle{apj}

\end{document}